\documentclass[reprint,amsmath,amssymb,pra,superscriptaddress,twocolumn,aps]{revtex4-2}

\usepackage{dcolumn}% Align table columns on decimal point
\usepackage{bm}% bold math
\usepackage{dsfont}
\usepackage{graphicx,epsfig}
\usepackage{amsmath}
\usepackage{braket}
\usepackage{caption}
\usepackage{float}
\usepackage[shortlabels]{enumitem}
\usepackage{hyperref}
\usepackage{blindtext}
\usepackage{color}
\usepackage[bottom]{footmisc}

\DeclareMathOperator{\Tr}{Tr}

\renewcommand{\eqref}[1]{Eq.~(\ref{#1})}
\newcommand{\figref}[1]{Fig.~\ref{#1}}
\newcommand{\appref}[1]{App.~\ref{#1}}
\newcommand{\secref}[1]{Sec.~\ref{#1}}

\begin{document}

\title{A contextuality witness inspired by optimal state discrimination}

\author{Carles Roch i Carceller}\email{crica@dtu.dk}\address{Center for Macroscopic Quantum States bigQ, Department of Physics, Technical University of Denmark, Fysikvej 307, 2800 Kgs. Lyngby, Denmark}

\author{Jonatan Bohr Brask}\address{Center for Macroscopic Quantum States bigQ, Department of Physics, Technical University of Denmark, Fysikvej 307, 2800 Kgs. Lyngby, Denmark}

\begin{abstract}
Many protocols and tasks in quantum information science rely inherently on the fundamental notion of contextuality to provide advantages over their classical counterparts, and contextuality represents one of the main differences between quantum and classical physics. In this work we present a witness for preparation contextuality inspired by optimal two-state discrimination. The main idea is based on finding the accessible averaged success and error probabilities in both classical and quantum models. We can then construct a noncontextuality inequality and associated witness which we find to be robust against depolarising noise and loss in the form of inconclusive events.
\end{abstract}

\maketitle

\section{Introduction}

In classical physics, properties of physical objects can be assumed to exist independently of any observation. However, quantum mechanics shows that attributes of physical systems do not exist predeterminedly,
 in the sense that it is not generally possible to consistently assign values to measurable quantities that are independent of which other quantities are jointly measured. This impossibility of reproducing the predictions of quantum mechanics with models that assign values independent of the measurement context is known as quantum contextuality \cite{budroni2022}.
 %This introduces the idea of quantum contextuality, a fundamental result that highlights the conflict between classical and quantum mechanics. It states that measurement outcomes in quantum mechanics do depend on the performance of other compatible measurements, in contrast with the independence predicted in classical models \cite{budroni2022}. Quantum mechanics states that the results of compatible measurements on any physical system depend on the context in which they are performed, i.e. what other properties are measured on it.
This concept originates with the Bell-Kochen-Specker theorem \cite{kochen1968,bell1966}, which demonstrates that quantum theory is incompatible with noncontextual hidden-variable models. It has been demonstrated that contextuality constitutes a resource for various applications in quantum information including magic states \cite{howard2014}, quantum key distribution \cite{bechmann2000}, device-independent security \cite{horodecki2010} and quantum randomness certification \cite{abbott2012,chailloux2016}. The traditional definition of contextuality requires a composite system, and its standard proof applies to Hilbert spaces of dimension three or higher \cite{klyachko2008,cabello2013}. The notion of (non)contextualilty has been further generalised in Ref.~\cite{spekkens2005}, based on operational equivalences and ontological models. Similarly to Kochen-Specker, generalised contextuality has also been proven to provide a resource for certain quantum information tasks. For instance parity-oblivious multiplexing \cite{spekkens2009,ghorai2018}, random-access codes \cite{ambainis2019}, quantum randomness certification \cite{carceller2022}, communication \cite{hameedi2017,debashis2019,saha2019}, and state discrimination \cite{schmid2018,flatt2022}. Quantum theory has also been shown to be less preparation contextual than the general operational theory known as box world \cite{banik2015}.

%\car{Similarly to Kochen-Specker, generalised contextuality has also been proven to provide a resource for certain quantum information tasks. For instance parity-oblivious multiplexing \cite{spekkens2009,ghorai2018}, random-access codes \cite{ambainis2019}, quantum randomness certification \cite{carceller2022}, communication \cite{hameedi2017,debashis2019,saha2019}, and state discrimination \cite{schmid2018,flatt2022}. Quantum theory has also been shown to be less preparation contextual than the general operational theory known as box world \cite{banik2015}.} 

In this work we aim to find a simple witness for generalised contextuality in the sense introduced in Ref.~\cite{spekkens2005}. While a number of contextuality witnesses exist in the literature \cite{krishna2017,kunjwal2015,kunjwal2018,mazurek2016,pusey2018,schmid2018ineq}, here we benefit from the simplicity of prepare-and-measure scenarios with two preparations and a single measurement. Such scenarios are of wide importance for both fundamental studies of quantum mechanics and applications in quantum technology including sensing, communication, and randomness generation \cite{bae2015,barnett2002,gomez2022}. We then find a contextuality witness in this framework, which is inspired by optimal two-state discrimination. Our results show that contextuality can be witnessed in the presence of both significant depolarising noise and loss. %We also consider loss, in addition to noise, in the form of inconclusive results. Although counterintuitive, in some cases, these appear to mitigate the witnessing worsening under the effects of white noise.

The rest of the paper is organised as follows. In \secref{sec:state_discrimination} we give a brief introduction to the basic notions in state discrimination in a theory-independent manner. We continue in \secref{sec:scenario} presenting the prepare-and-measure scenario and the goal that defines the main state discrimination task in order to properly define the witness independently in both quantum and noncontextual models. Finally, we discuss the main results of the paper in \secref{sec:discussion} and conclude the work in \secref{sec:conclusion}.

\section{Basic notions in state discrimination}
\label{sec:state_discrimination}

Any state discrimination scenario is formed by state preparations and effects \cite{bae2016,kimura2009}. The former are labeled by preparation procedures $x\in X$ and the latter by the possible answers $b\in B$ to the questions in $X$. The gathered data is usually expressed as conditional probabilities (correlations) $p(b|x)$. The goal in state discrimination is to determine $x$ from the transmitted states, i.e. to achieve $b=x$. For any particular model (e.g.\ quantum or noncontextual), an optimisation problem can be built obeying the constraints of that model. As is customarily done in state discrimination settings, we denote the probability $p(b=x|x)$ of a correct answer the \textit{success} probability, whereas $p(b\neq x|x)$ for $b \in X$ is called the \textit{error} probability. One must also consider events where the answer $b$ is not in the set of questions $X$ (i.e. $X\subset B$). We group answers not in $X$ and label them by $b=ø$. We denote $p(b=ø|x)$ the \textit{inconclusive} probability. 

Success, error, and inconclusive probabilities each play a different role in the discrimination scenario \cite{barnett2009,bae2015,bergou2007}. Different state discrimination tasks can be defined by different figures of merits, which are functions of the observed conditional probabilities, and different constraints on these probabilities. For example, the goal in minimum-error state discrimination (MESD) is to maximise the success probability whilst inconclusive events do not occur \cite{loubenets2022,bae2013,herzog2004} (hence converting the goal into a minimisation of the error probability due to normalisation). On the other hand, in unambiguous state discrimination (USD), the goal is also to maximise the success probability, with the main constraint that error probabilities must vanish \cite{ivanovic1987, dieks1988, peres1988} (thus converting the goal into a minimisation of inconclusive probabilities). Lastly, in maximum-confidence state discrimination (MCSD), the goal is to maximise the confidence $C$, i.e. the probability of receiving input $x$ given the outcome $b=x$, which can be expressed as the success probability divided by the rate of events of interest \cite{jimenez2011,croke2006,herzog2009,herzog2012,herzog2015,bagan2012}. Concretely $C := p_x p(x|x)/\eta_x$, for $\eta_b=\sum_x p_x p(b|x)$, where $p_x$ are the prior probabilities for each preparation $x$. No further constraints are applied to MCSD, making it rather a more general approach. Also, it can be reduced to MESD and USD as particular cases. If $C=1$ the input $x$ must be unambiguously identified, resulting in USD, while MESD is recovered by adopting $\sum_x \eta_x C_x$ as the figure of merit. 

\section{Scenario}
\label{sec:scenario}

In the following, we focus on two-state discrimination, characterised by the sets of preparations $X=\left\{0,1\right\}$, considered equiprobable, and outcomes $B=\left\{0,1,ø\right\}$. We also introduce the averaged success $p_{\text{suc}}$, error $p_{\text{err}}$, and inconclusive $p_{\text{inc}}$ probabilities as
\begin{align}
\label{eq:sqeq}
    p_{\text{suc}} &:= \frac{1}{2}\left(p(0|0)+p(1|1)\right) , \\
    p_{\text{err}} &:= \frac{1}{2}\left(p(1|0)+p(0|1)\right) , \\
    p_{\text{inc}} &:= \frac{1}{2}\left(p(ø|0)+p(ø|1)\right) = 1 - p_{\text{suc}} - p_{\text{err}} .
\end{align}
We will fix $p_{\text{inc}}$ and ask the following question: which regions in correlation space, parameterized by $p_{\text{suc}}$ and $p_{\text{err}}$, are feasible in quantum mechanics or in a noncontextual model? The answer to this question is not trivial if state preparations are not perfectly distinguishable. For fixed inconclusive rate, the sum $p_{\text{suc}}+p_{\text{err}} = 1 - p_{\text{inc}}$ is fixed and we can focus on the difference. We therefore define the following witness on the level of probabilities
\begin{align}
    \label{eq:witness_probs}
    \mathcal{W} :=  \frac{1}{2}\left(p_{\text{suc}}- p_{\text{err}}\right) \ .
\end{align}
For each model, we will separately formulate an  optimisation problem and find a bound on $\mathcal{W}$. The feasible region is necessarily convex since, for two different measurement strategies producing different behaviours, probabilistically choosing between them (using local randomness) defines another valid measurement strategy. The corresponding behaviour will then be the convex combination of the first two behaviours. We can thus use techniques in convex optimisation to efficiently solve the maximisation problem for each model.

\subsection{Quantum model}
\label{subsec:quantum_model}

Consider an ensemble of two noisy states $\rho_x=r_{s}\ket{\psi_x}\bra{\psi_x}+(1-r_{s})\mathds{1}/2$ for $x=0,1$, with distinguishability characterised by the overlap $\delta=|\braket{\psi_0|\psi_1}|$. Let $\hat{\pi}_b$ represent a valid POVM for $b=0,1,ø$, such that $\Tr\left[\rho_x\hat{\pi}_b\right]=p(b|x)$. Our goal is to find the maximum difference between success and error probabilities, for a fixed inconclusive rate. To do so, let us introduce the following operator 
\begin{align}
\label{eq:WQ}
    \hat{\Delta}_{x} := \frac{(-1)^{x}}{2}\left(\hat{\pi}_0 - \hat{\pi}_1\right) \ .
\end{align}
We must find the maximum difference 
\begin{align}
\label{eq:DeltaQ}
    \mathcal{W}^{\text{Q}} := & \max \frac{1}{2}\left(p^{\text{Q}}_{\text{suc}} - p^{\text{Q}}_{\text{err}}\right) = \max \sum_{x}\Tr\left[\hat{\Delta}_x \rho_x\right] \ ,
\end{align}
where the optimisation is over all measurements forming valid POVMs, $\hat{\pi}_b \geq 0$ and $\sum_{b}\hat{\pi}_b = \mathds{1}$, and subject to $p_{\text{inc}}=\frac{1}{2}\Tr\left[(\rho_0+\rho_1)\hat{\pi}_{ø}\right]$. This maximisation can be rendered as a semi-definite program (SDP) \cite{vandenberghe1996}.

In \appref{app:optimal_measurements} we find an analytical form of the optimal measurement. The solution to \eqref{eq:DeltaQ} is given by
\begin{align}
    &\mathcal{W}^{\text{Q}} = \frac{r_{s}}{2}\sqrt{(1-\delta^2)(1-\frac{2p_{\text{inc}}}{1+r_{s}\delta})}  && \text{for} \ p_{\text{inc}}\leq r_{s}\delta \\
    &\mathcal{W}^{\text{Q}} = \frac{r_{s}}{2}\sqrt{1-\delta^2}\sqrt{1-r_{s}^{2}\delta^{2}}\frac{1-p_{\text{inc}}}{1-r_{s}^{2}\delta^{2}} && \text{for} \ p_{\text{inc}}\geq r_{s}\delta \nonumber \ .
\end{align}
One can also write down the optimal success and error probabilities as
\begin{align}
\label{eq:sqeq2}
    p_{\text{suc}}^{\text{Q}} &= \frac{1}{2}\left(1+2\mathcal{W}^{\text{Q}}-p_{\text{inc}}\right) \ ,
\end{align}
and $p_{\text{err}}^{\text{Q}}=1-p_{\text{suc}}^{\text{Q}}-p_{\text{inc}}$. 

The success and error probabilities we found are the maximal and minimal probabilities according to quantum theory in a qubit state-discrimination problem. Interestingly, one can recover the bounds from other protocols as specific cases. For instance, if the experiment only produces conclusive outcomes ($p_{\text{inc}}=0$), the problem is reduced to the usual MESD. Then, one recovers the Helstrom bound as a minimum error rate $p_{\text{err}}$ \cite{bae2013,helstrom1968,qiu2008}. On the other hand, if the experiment is designed with a null error rate ($p_{\text{err}}=0$) and zero noise ($r_{s}=1$), one recovers USD. In that case, the maximal success probability is $p_{\text{suc}}=1-\delta$,  leaving a minimal rate of inconclusive events $p_{\text{inc}}=\delta$, the minimal value for USD \cite{kleinmann2010,salazar2012}. Finally, one can directly compute the maximum confidence of the whole ensemble by writing $C=p_{\text{suc}}/(p_{\text{suc}}+p_{\text{err}})$. One recovers the maximum confidence obtained in \cite{flatt2022} if $p_{\text{inc}}\leq r_{s}\delta$. For larger values of the inconclusive rate, one can still compute the maximum confidence with the same formula since MCSD and the present scheme share the exact same goal (maximise the success and minimize the error probabilities).

\begin{figure}
    \centering
    \hspace{-0.5cm}\includegraphics[width=0.49\textwidth]{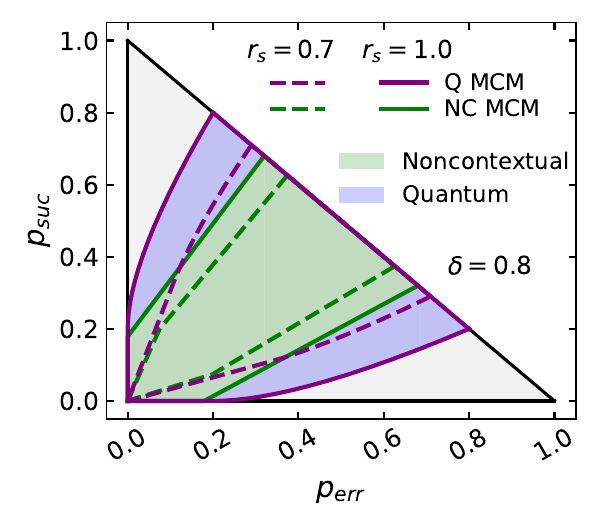}
    \vspace{-0.5cm}
    \caption{Space of probabilities corresponding to a two-state discrimination setting. Continuous lines denote maximum confidence measurements in both quantum (purple) and noncontextual (green) models. Even with a bounded value of noise ($r_{s}=0.7$), the MCM line according to the quantum model falls outside the noncontextual region.}
    \label{fig:pspace_quantum}
\end{figure}

\subsection{Noncontextual model}
\label{subsec:noncontextual_model}

We now outline a non-contextual ontological model for the prepare-and-measure scenario \cite{spekkens2005,spekkens2008,schmid2018}. The system is associated with an ontic state space $\Omega$ in which each point $\lambda$ completely defines all physical properties, i.e.~the outcomes of all possible measurements. Each state preparation $x$ samples the ontic state space according to a probability distribution $\mu_{x} (\lambda)$, referred to as the \textit{epistemic state}. Each measurement is defined by a set of \textit{response functions}, that is, non-negative functions $\xi_b(\lambda)$ over the ontic space, such that $\sum_b \xi_b(\lambda) = 1$ for all $\lambda \in \Omega$. The probability of obtaining the outcome $b$ when state $\mu_x$ was prepared is then
\begin{equation}
\label{eq.ontprob}
p(b|x) = \int_\Omega d\lambda \  \mu_x(\lambda) \xi_b(\lambda) .
\end{equation}
While distinct ontic states can be perfectly discriminated, epistemic states with overlapping distributions cannot. Distinguishability can be quantified in terms of the \textit{confusability} between two epistemic states $\mu_{x}$ and $\mu_{y}$:
\begin{align}
\label{eq:confusability}
    c_{xy} := \int_{\text{supp}\left[\mu_x(\lambda)\right]} d\lambda \ \mu_y(\lambda)
\end{align}
It is the discrimination of epistemic states which we compare against quantum state discrimination. 

\begin{figure*}
    \centering
    \includegraphics[width=1.0\textwidth]{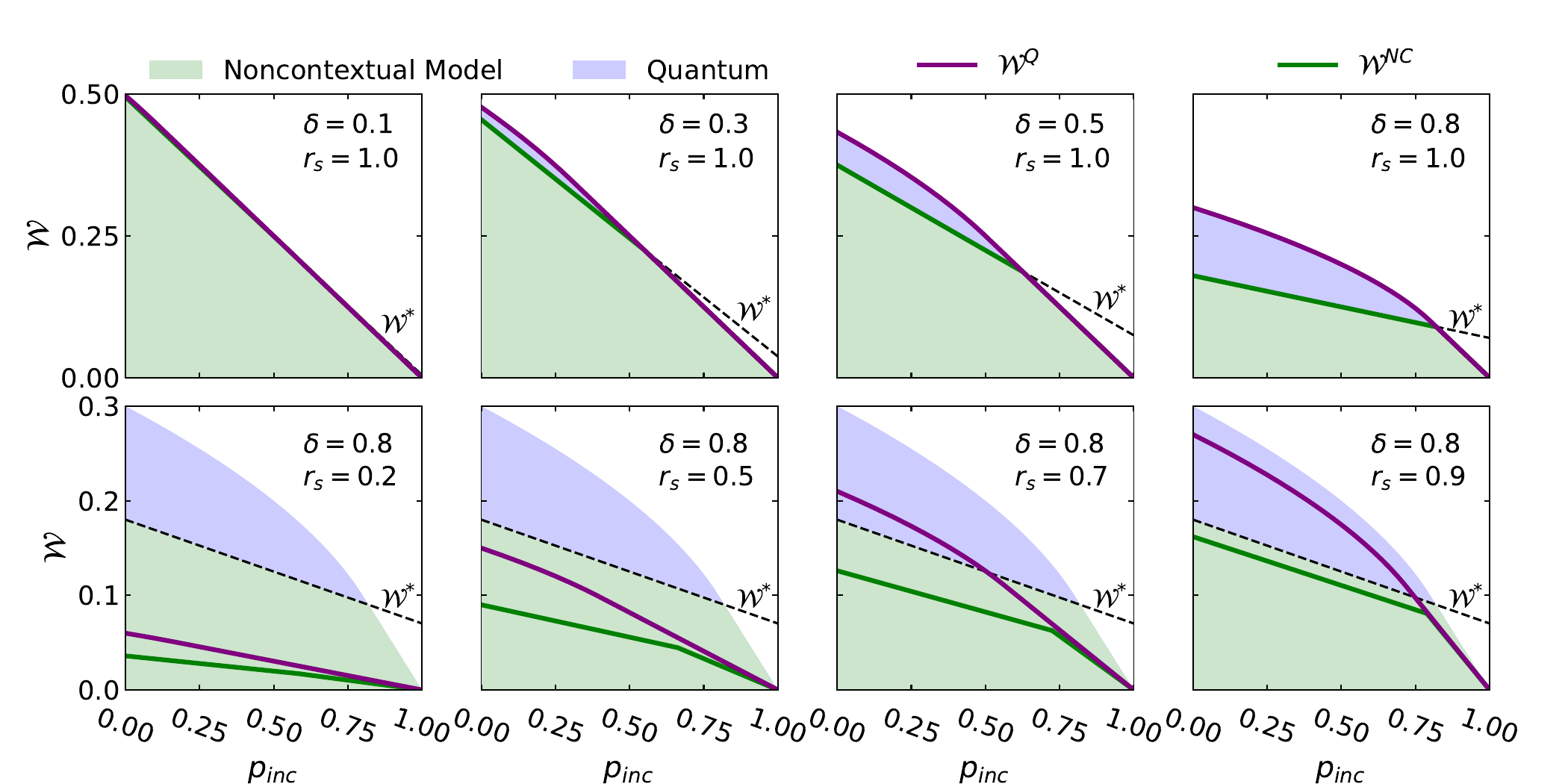}
    \vspace{-0.5cm}
    \caption{Bounds on the witness $\mathcal{W}$ according to quantum and noncontextual models. On the first row we show noiseless cases with different overlaps. Below, on the second row, we fix a particular overlap and show the effects of depolarising noise on the preparation. The green area denotes the feasible values according to quantum and noncontextual models and the blue region solely for the quantum model. The black-dashed line shows the contextuality witness $\mathcal{W}^{\ast}$ in \eqref{eq:NCineq}. Any behavior above $\mathcal{W}^{\ast}$ is an evidence of contextuality.}
    \vspace{-0.5cm}
    \label{fig:Delta_comp}
\end{figure*}

Furthermore, we require the ontological model to be preparation-noncontextual. Two preparations are said to be operationally equivalent if they cannot be distinguished by any measurement, and an ontological model is said to be preparation-noncontextual if all operationally equivalent preparations are assigned to the same epistemic state. To impose noncontextuality on the ontological model, we assume the existence of a particular pair of pure states $S:=\left\{\mu_{0},\mu_{1}\right\}$ and complementary states $S^{\perp}:=\left\{\mu_{0}^{\perp},\mu_{1}^{\perp}\right\}$, i.e.\ $\mu_x$ and $\mu_x^{\perp}$ have non-overlapping supports $\mu_x(\lambda)\mu_x^\perp(\lambda) = 0 \,\, \forall \, \lambda$. The pairwise confusability $c_{01}$ of $\mu_0$ and $\mu_1$ is the same as for $\mu_0^\perp$ and $\mu_1^\perp$. Preparation noncontextuality implies that preparing $\mu_x$ and $\mu_x^{\perp}$ with equal probability for $x=0$ or $x=1$ must be equivalent \cite{schmid2018,spekkens2005}, that is $\frac{1}{2}\mu_0 + \frac{1}{2}\mu_0^{\perp} = \frac{1}{2}\mu_1 + \frac{1}{2}\mu_1^{\perp}$. This statement implies that any pair from the set of states $\{\mu_0,\mu_1,\mu_0^\perp,\mu_1^\perp\}$ are equal on their overlap, i.e. 
\begin{align}
    \mu_{0}(\lambda) = \mu_{1}(\lambda) \ \forall \lambda \in \text{supp}\left[\mu_{0}(\lambda)\right]\cup\text{supp}\left[\mu_{1}(\lambda)\right] \, \forall \, \lambda , 
\end{align}
and similarly for the other pairs. This, in turn, results in symmetric confusabilities, $c_{01}=c_{10}:=c$. Quantum and noncontextual models can be then compared through $\delta^{2}=c$. The noncontextual model we use in this work can be understood as an attempt to describe quantum theory, and in general it will reproduce some quantum correlations but not all, as we explore below.

We now present the main problem in a noncontextual model. The two preparations are represented by the following epistemic states affected by depolarising noise
\begin{align}
    \tilde{\mu}_{0}(\lambda) &= r_{s}\mu_{0}(\lambda) + (1-r_{s})\mu_{\mathds{1}/2}(\lambda) \\
    \tilde{\mu}_{1}(\lambda) &= r_{s}\mu_{1}(\lambda) + (1-r_{s})\mu_{\mathds{1}/2}(\lambda) \nonumber
\end{align}
These can be characterised by the confusability of the noiseless epistemic states $c:=c_{10}$ from \eqref{eq:confusability}. We also consider a single measurement with two conclusive outcomes $b=0,1$ and an inconclusive result $b=ø$, represented by the response functions $\xi_{b}(\lambda)$. Let us define the analogous observable to \eqref{eq:WQ}
\begin{align}
    \Delta_{x}^{\text{NC}}(\lambda) := \frac{(-1)^{x}}{2}\left(\xi_0(\lambda)-\xi_{1}(\lambda)\right) \ .
\end{align}
Then, we can rewrite the problem as a maximisation
\begin{align}
\label{eq:DeltaNC}
    \mathcal{W}^{\text{NC}} :=& \max \frac{1}{2}\left(p^{\text{NC}}_{\text{suc}} - p^{\text{NC}}_{\text{err}}\right) \\
    =& \max \sum_{x}\int_{\Omega} d\lambda \ \tilde{\mu}_{x}(\lambda) \Delta_{x}^{\text{NC}}(\lambda) \ , \nonumber
\end{align}
subject to $\xi_{b}(\lambda)$ being valid response functions, $\xi_{b}(\lambda)\geq 0$ and $\sum_{b}\xi_{b}(\lambda)=1$, $\forall \lambda$, and a given rate of inconclusive events $p_{\text{inc}}= \frac{1}{2}\int d\lambda \left(\tilde{\mu}_{0}(\lambda)+\tilde{\mu}_1(\lambda)\right)\xi_{ø}(\lambda)$. In \appref{app:optimal_measurements} we show how this maximisation can be rendered as a simple linear problem, for which we are able to find an analytical solution
\begin{align}
    \mathcal{W}^{\text{NC}} &= \frac{r_{s}}{2}\left(1-c\right)\left(1-\frac{p_{\text{inc}}}{1+r_{s}c}\right) \quad \text{for} \ p_{\text{inc}}\leq (1+r_{s}c)/2 \nonumber \\
    \mathcal{W}^{\text{NC}} &= \frac{1}{2}(1-p_{\text{inc}})\frac{r_{s}(1-c)}{1-r_{s}c}  \quad \text{for} \ p_{\text{inc}}\geq (1+r_{s}c)/2 \ .
\end{align}
This results in the following success and error probabilities. For $p_{\text{inc}}\leq (1+r_{s}c)/2$
\begin{align}
\label{eq:snc_enc_low}
    p^{\text{NC}}_{\text{suc}} &= \frac{1+r_{s}}{2}\left(1-\frac{p_{\text{inc}}}{1+r_{s}c}\right) - \frac{r_{s}c}{2} \ ,
\end{align}
and for $p_{\text{inc}}\geq (1+r_{s}c)/2$
\begin{align}
\label{eq:snc_enc}
    p^{\text{NC}}_{\text{suc}} &= \frac{1}{2}\left(1+2\mathcal{W}^{\text{NC}}-p_{\text{inc}}\right) \ ,
\end{align}
and $p_{\text{err}}^{\text{NC}}=1-p_{\text{suc}}^{\text{NC}}-p_{\text{inc}}$. 

\section{Discussion}
\label{sec:discussion}

In \figref{fig:pspace_quantum} we show the achievable probabilities in quantum and noncontextual models. The white region delimited by the black contour shows the feasible space in the case of fully distinguishable preparations. That is, when states can be directly identified with ontic states $\lambda$. The area shaded in blue shows the feasible space according to quantum theory.  In its contour we find $p_{\text{suc}}^{\text{Q}}$ from \eqref{eq:sqeq}, for $r_{s}=1$. The region reproducible by the noncontextual model (green area) is contained in the quantum set. Similarly, we find $p_{\text{suc}}^{\text{NC}}$, from \eqref{eq:snc_enc_low}, in its contour, also for $r_{s}=1$. We see that the quantum predictions depart from classical (noncontextual) interpretations. Increasing the overlap $\delta=\sqrt{c}$, this distinction becomes more pronounced, and at the same time both the quantum and noncontextual feasible spaces shrink.
%Both the quantum and noncontextual feasible space shrink with increasing overlap $\delta=\sqrt{c}$. We see that the quantum predictions depart from classical (noncontextual) interpretations.

Moreover, we can identify some extremes of the quantum region with the bounds found in each state discrimination protocol. The diagonal line that delimits the upper-right part of the feasible regions covers the state discrimination scenarios with zero inconclusive rates. The vertices of the quantum region on that line reproduce the Helstrom bound \cite{helstrom1967,helstrom1968,helstrom1969} obtained in MESD. The same applies for the vertices corresponding to the noncontextual line, which reproduce the maximal success probabilities in MESD obtained in \cite{schmid2018}. Also, the maximal $p_{\text{err}}$ and $p_{\text{suc}}$ on the flat part of the bottom and left-most boundary, respectively, reproduce the maximal unambiguous error and success rates for quantum \cite{ivanovic1987} and noncontextual \cite{flatt2022} models, obtained in USD. Finally, the entire quantum boundary (purple line) corresponds to a maximum confidence measurement (MCM)\cite{croke2006,lee2022} for the ensemble of qubit states (i.e.\ maximising the average confidence). MCM thus provides optimal success probability in any (qubit-)state discrimination scenario. Similarly, the noncontextual boundary is obtained by a noncontextual MCM. We can see, by writing $C=p_{\text{suc}}/(p_{\text{suc}}+p_{\text{err}})$, that the confidence coincides with the bounds found in the literature \cite{croke2006,herzog2009,jimenez2011,lee2022,flatt2022}.

When depolarizing noise is included, the bounds on all protocols depart from the borders of the quantum region. The Helstrom bound from both quantum and noncontextual MESD comes closer to the center of the probability space as noise increases. Also, when noise is taken into account, USD is not possible as here we can see that the bottom and left-most borders are not reachable. The space enclosed by the MCM lines also narrows. Indeed, noise makes the prepared physical states less distinguishable in both models. For a given noisy ensemble, the points on the quantum region outside the MCM lines are not accessible. 

\begin{figure}
    \centering
    \includegraphics[width=0.5\textwidth]{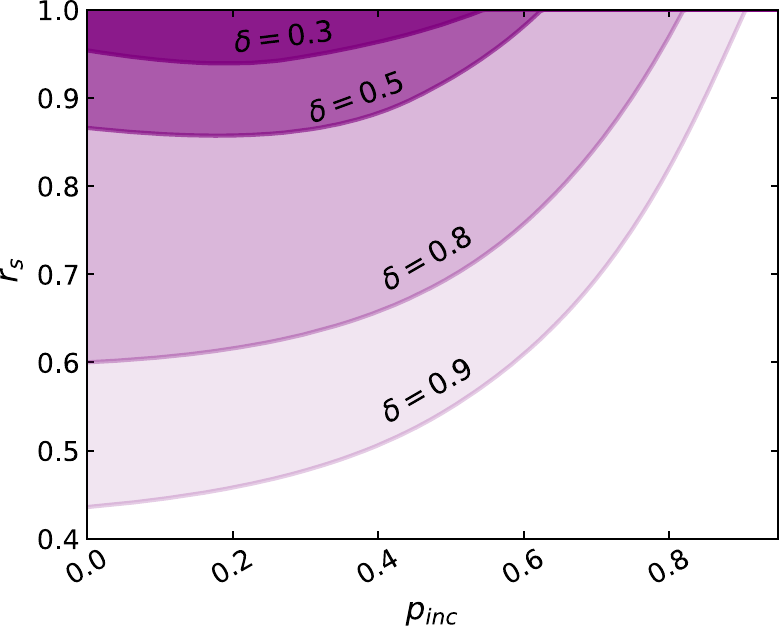}
    \caption{Tolerable amount of depolarising noise for which our witness can detect quantum contextuality as function of inconclusive rate $p_\text{inc}$ for different overlaps $\delta$. Contextuality is witnessed in the shaded regions above the solid lines (larger $r_s$ means less noise).}
    \label{fig:r_min}
\end{figure}

A different perspective is plotted in \figref{fig:Delta_comp}. Contextual behavior is manifested in the blue shaded region above the dashed black line, which corresponds to the inequality 
\begin{equation}
\label{eq:NCineq}
    \mathcal{W} \leq \mathcal{W}^{*} = \left. \mathcal{W}^\text{NC}\right|_{r_s=1} = \frac{1-\delta^{2}}{2}\left(1-\frac{p_{\text{inc}}}{1+\delta^{2}}\right) \ ,
\end{equation}
for noiseless preparations with distinguishability bounded by overlap $\delta$ (quantum) or confusability $\delta^2$ (noncontextual). Here, $\mathcal{W}^{*}$ is the noncontextual bound without noise. Note that it is sufficient to lower-bound the confusability (or equivalently the overlap) because our bounds on $\mathcal{W}$ decrease as preparations become less distinguishable. Also, note that the witness places no assumptions on the measurement, which is completely uncharacterised. We see, from the two lower-right plots in \figref{fig:Delta_comp} that noncontextuality can be witnessed in the presence of fairly high values of noise (for example, $r_s=0.7$ equivalent to $30 \%$ depolarising noise).

We finally look at the amount of depolarising noise that our witness can tolerate while still being able to detect contextuality. Depolarising noise in the preparation is parameterized through $r_s$ (see above \eqref{eq:WQ}). In \figref{fig:r_min} we show the minimum tolerable $r_s$ for which contextuality can be witnessed, i.e.\ for which $\mathcal{W}^{\rm Q}\geq \mathcal{W}^{*}$. Two noiseless preparations with confusability $c=\delta^{2}$ in a noncontextual model can reproduce all quantum correlations from a two-state discrimination scenario with fixed $p_{\rm inc}$, as long as $r_{s}$ is below the plotted lines. Noise tolerance is higher for larger confusabilities. The cases with null inconclusive outcomes are covered in Refs.\cite{schmid2018,vinicius2023}. Remarkably, observe that for low noise, the appearance of inconclusive events strengthens robustness, as is e.g.\ the case for $\delta=0.3$ or $\delta=0.5$ in \figref{fig:r_min}. An intuitive explanation for this might be that, in the absence of noise, the optimal rate of inconclusive events is $p_{\text{inc}}= \delta$ (USD is impossible for lower $p_{inc}$). With noise however, the error probability cannot be zero, which allows for lower optimal inconclusive probabilities. This is reflected in \figref{fig:r_min}, where the minima in each curve is given by an inconclusive probability lower than $\delta$.  \\

\section{Conclusion}
\label{sec:conclusion}

In this work, we presented a witness of contextuality in two-state discrimination scenarios. We started by formulating the problem of finding the optimal measurement in two-state discrimination settings. We consider a measurement to be optimal if it maximises the difference between success and error probabilities. This leads to correlations reaching the boundary of the feasible space parameterized by success and error probabilities. Parametrizing the correlation space in this manner allows us to clearly distinguish the feasible sets for the quantum and noncontextual models. Additionally, allowing for inconclusive events we found that maximum-confidence measurements are optimal in both quantum and noncontextual models. That led to the definition of the witness $\mathcal{W}$ in \eqref{eq:witness_probs} and the inequality \eqref{eq:NCineq}. This inequality allows for a flexible rate of inconclusive events (e.g.\ due to losses) and is robust against depolarising noise, as we show in \figref{fig:r_min}. Even more importantly, our results show that in some cases incorporating inconclusive results can help strengthening the noise tolerance in terms of witnessing quantum contextuality. Thus, our results open new avenues for exploring contextual advantages in realistic scenarios, using inconclusive results as a benefit towards noise robustness.

The results we present in this work are explicitly derived for two state discrimination scenarios. However, we strongly believe they can be generalised to multiple state discrimination scenarios. Although we left this as an open question out of the scope of this work, intuitively, it can be done in a twofold manner, depending on the goal in the discrimination task. If one is interested in discriminating all states equally, the problem can be reduced to pairwise state discrimination sub-tasks. Hence reducing an $N$ state to an individual two-state discrimination scenario as presented in this work. Otherwise, if one aims to distinguish one state from the ensemble, the problem can again be reduced to the discrimination between the state of interest and the mixture of the rest of the ensemble. This is equivalent to our scenario, but only one of the states is affected by depolarising noise. Thus, at the end of the day, one can think of more general cases and find our results still applicable. \\

We gratefully acknowledge support from the Danish National Research Foundation, Center for Macroscopic Quantum States (bigQ, DNRF142), VILLUM FONDEN (research grant 40864), the Carlsberg Foundation CF19-0313, and a DTU-KAIST
Alliance Stipend.

\bibliography{apssamp}% Produces the bibliography via BibTeX. 

\begin{widetext}

\newpage
    
\appendix

\section{Optimal measurements}
\label{app:optimal_measurements}

In the supplemental material, we derive the optimal measurements for both the quantum and noncontextual models considered in the main text. We call a measurement optimal when it maximises the probability success whilst the error probability is kept at minimum.

\subsection{Quantum model}

Let us start by writing the semi-definite program that finds the optimal measurement according to the quantum theory. We express the difference we aim to maximise as
\begin{align}
     \mathcal{W}^{\text{Q}}=\frac{1}{2}\left(p_{\text{suc}}^{\text{Q}}-p_{\text{err}}^{\text{Q}}\right) &= \frac{1}{2}\left(\Tr\left[\rho_0\hat{\pi}_0\right]+\Tr\left[\rho_1\hat{\pi}_1\right]\right) - \frac{1}{2}\left(\Tr\left[\rho_0\hat{\pi}_1\right]+\Tr\left[\rho_1\hat{\pi}_0\right]\right) \nonumber \\
     &= \frac{1}{2}\Tr\left[(\rho_0-\rho_1)(\hat{\pi}_0-\hat{\pi}_1)\right] = \sum_{x=0}^{1}\Tr\left[\rho_x \hat{\Delta}_{x}\right] \ ,
\end{align}
where we included the operator $\hat{\Delta}_{x}$ introduced in the main text. One can write down the problem in the following SDP form:
\begin{align}
\label{eq:SDPlimits}
    \underset{\left\{\hat{\pi}_b\right\}}{\text{maximise}} & \quad \frac{1}{2}\Tr\left[(\rho_0-\rho_1)(\hat{\pi}_0-\hat{\pi}_1)\right] \nonumber \\
    \text{subject \ to:} & \quad \hat{\pi}_b \geq 0 \quad \sum_{b}\hat{\pi}_b = \mathds{1}  \\
    & \quad \frac{1}{2}\Tr\left[(\rho_0+\rho_1)\hat{\pi}_{ø}\right] = p_{\text{inc}} \nonumber
\end{align}
We begin by analysing the optimality conditions of the POVM $\hat{\pi}_b$. The corresponding Lagrangian is given by
\begin{align}
    \mathcal{L} &= \frac{1}{2}\Tr\left[(\rho_0-\rho_1)(\hat{\pi}_0-\hat{\pi}_1)\right] + \sum_{b}r_b\Tr\left[\hat{\pi}_b\sigma_b\right] + \Tr\left[K\left(\mathds{1}-\sum_{b}\hat{\pi}_b\right)\right] + s\left(p_{\text{inc}}-\frac{1}{2}\Tr\left[(\rho_0+\rho_1)\hat{\pi}_{ø}\right]\right) \ ,
\end{align}
where we introduced the following dual variables: $r_b\sigma_b$ for the PSD constraint, $K$ accounting for the normalisation constraint and $s$ for the constraint fixing the inconclusive rate. The dual problem can be straight formulated through the supremum of the Lagrangian
\begin{align}
\label{eq:supplims}
    \mathcal{S} = \underset{\hat{\pi}_b}{\text{supp}} \ \Tr\left[K\right] + sp_{\text{inc}} + \sum_{b}\Tr\left[\hat{\pi}_b G_b\right] \ ,
\end{align}
where we defined
\begin{align}
    G_b &= \frac{1}{2}\Tr\left[\left(\rho_0-\rho_1\right)\left(\delta_{b,0}-\delta_{b,1}\right)\right] + r_b\sigma_b - \frac{1}{2}\left(\rho_0+\rho_1\right)\delta_{b,ø}-K \ .
\end{align}
The supremum in \eqref{eq:supplims} will diverge unless $G_b=0$. This leads to the Lagrange Stability optimality condition. Complementary slackness reads $r_b\Tr\left[\sigma_b\hat{\pi}_b\right]=0$, which in a qubit space means that $\pi_b$ must be rank-1. With all that, we are ready to derive the form of the optimal POVM. We consider a pair of pure states $\rho_x$ oriented symmetrically with respect to the $Z$ pole in the Bloch sphere. That means
\begin{align}
 \label{eq:rholimits}
    \rho_0 = \frac{1}{2}\left[\mathds{1}+r\sqrt{1-\delta^{2}}X + r_{s}\delta Z\right] \quad , \qquad \rho_1 = \frac{1}{2}\left[\mathds{1}-r_{s}\sqrt{1-\delta^{2}}X + r\delta Z\right]  \ .
\end{align}
By fixing the orientation of the states in that way, the problem acquires a symmetry with respect to the states. This is very convenient since we can directly find the analytical form of $\hat{\pi}_{ø}$ through the last constraint in \eqref{eq:SDPlimits}: $\hat{\pi}_{ø}=\frac{p_{\text{inc}}}{1+r_{s}\delta}\left[\mathds{1}+Z\right]$. The other two POVM elements ($\hat{\pi}_0$ and $\hat{\pi}_1$) can be also directly found noting that the maximum in \eqref{eq:SDPlimits} is reached when $\hat{\pi}_0-\hat{\pi}_1$ is proportional to $\rho_0-\rho_1=r_{s}\sqrt{1-\delta^{2}}X$ (according to \eqref{eq:rholimits}). At the end of the day, this leaves us with the following optimal POVM:

\begin{align}
\label{eq:POVMlimits}
    \hat{\pi}_0 =& \frac{1}{2}\left(1-\frac{p_{\text{inc}}}{1+r_{s}\delta}\right)\left[\mathds{1}+\frac{1+r_{s}\delta}{r_{s}\sqrt{1-\delta^2}}\frac{\mathcal{W}^{\text{Q}}}{1+r_{s}\delta-p_{\text{inc}}}X-\frac{p_{\text{inc}}}{1+r_{s}\delta-p_{\text{inc}}}Z\right] \\
    \hat{\pi}_1 =& \frac{1}{2}\left(1-\frac{p_{\text{inc}}}{1+r_{s}\delta}\right)\left[\mathds{1}-\frac{1+r_{s}\delta}{r_{s}\sqrt{1-\delta^2}}\frac{\mathcal{W}^{\text{Q}}}{1+r_{s}\delta-p_{\text{inc}}}X-\frac{p_{\text{inc}}}{1+r_{s}\delta-p_{\text{inc}}}Z\right] \\
    \hat{\pi}_{ø} =& \frac{2p_{\text{inc}}}{1+r_{s}\delta}\frac{1}{2}\left[\mathds{1}+Z\right]  \ ,
\end{align}
for $p_{\text{inc}}\leq r_{s}\delta$, and 
\begin{align}
\label{eq:POVMlimits2}
    \hat{\pi}_0 =& \frac{1}{2}\frac{1-p_{\text{inc}}}{1-r_{s}^2 \delta^2}\left[\mathds{1}+\sqrt{1-r_{s}^2\delta^2}X-r_{s}\delta Z\right] \\
    \hat{\pi}_1 =& \frac{1}{2}\frac{1-p_{\text{inc}}}{1-r_{s}^2 \delta^2}\left[\mathds{1}-\sqrt{1-r_{s}^2\delta^2}X-r_{s}\delta Z\right] \\
    \hat{\pi}_{ø} =& \frac{2(p_{\text{inc}}-r_{s}^2\delta^2)}{1-r_{s}^2\delta^2}\frac{1}{2}\left[\mathds{1}+\frac{1-p_{\text{inc}}}{p_{\text{inc}}-r_{s}^2 \delta^2} r_{s}\delta Z\right]  \ ,
\end{align}
for $p_{\text{inc}}\geq r_{s}\delta$. That POVM yields the optimal measurement that maximises the difference between success and error probabilities for a fixed rate of inconclusive events.

\subsection{Noncontextual model}

In a noncontextual model we can write the problem in the following maximisation form:
\begin{align}
\label{eq:sdpbelow}
    \underset{\xi_{b}(\lambda)}{\text{maximise}} & \quad \frac{1}{2}\int  d\lambda \left(\tilde{\mu}_{0}(\lambda)-\tilde{\mu}_1(\lambda)\right)\left(\xi_{0}(\lambda)-\xi_1(\lambda)\right) \\
    \text{subject to:} & \quad \xi_{b}(\lambda) \geq 0 , \ \sum_{b}\xi_{b}(\lambda) = 1 \ \forall \lambda \nonumber \\
    & \quad p_{\text{inc}} = \frac{1}{2}\int d\lambda \left(\tilde{\mu}_{0}(\lambda)+\tilde{\mu}_1(\lambda)\right)\xi_{ø}(\lambda) \ . \nonumber
\end{align}
We consider a pair of noisy epistemic states affected by depolarising noise:
\begin{align}
    \tilde{\mu}_{0}(\lambda) = r_{s}\mu_{0}(\lambda) + (1-r_{s})\mu_{\mathds{1}/2}(\lambda) \quad , \qquad \tilde{\mu}_{1}(\lambda) = r_{s}\mu_{1}(\lambda) + (1-r_{s})\mu_{\mathds{1}/2}(\lambda) \ .
\end{align}
These are characterised by the confusability of the noiseless states given by
\begin{align}
    c_{10} = \int_{\text{supp}\left[\mu_0(\lambda)\right]} d\lambda \mu_{1}(\lambda) \ .
\end{align}
For low enough rates of inconclusive events, the optimal response functions are those which unambiguously discriminate the noiseless epistemic states. These are of the following form
\begin{equation} \label{eq:respfuncbelow}
\xi_{0} (\lambda) = \begin{cases} q & \mbox{if } \lambda \in {\rm supp} [\overline{\mu}_{1}(\lambda)] \\ 0 & \mbox{otherwise } \ . \end{cases} \quad \xi_{1} (\lambda) = \begin{cases} q & \mbox{if } \lambda \in {\rm supp} [\overline{\mu}_{0}(\lambda)] \\ 0 & \mbox{otherwise } \ . \end{cases} 
\end{equation}
One can determine the value of $q$ in terms of the rate of inconclusive events and obtain
\begin{align}
\label{eq:qbelow}
    q = \frac{1-p_{\text{inc}}}{1-r_{s} c_{10}}
\end{align}
leaving the following extremal success and error probabilities:
\begin{align}
\label{eq:snc_enc}
    p_{\text{suc}}^{\text{NC}} = \frac{1-p_{\text{inc}}}{2}\left(1+\frac{r_{s}(1-c_{10})}{1-r_{s}c_{10}}\right) \quad p_{\text{err}}^{\text{NC}} = \frac{1-r_{s}}{2}\frac{1-p_{\text{inc}}}{1-r_{s} c_{10}} \ .
\end{align}
Normalisation implies in \eqref{eq:qbelow} that $p_{\text{inc}}\geq (1+r_{s}c_{10})/2$. In the noiseless case, that is the lower bound on the rate of inconclusive events in USD. Also, note that the confidence can be written as $C=p_{\text{suc}}/(1-p_{\text{inc}})$, which according to \eqref{eq:snc_enc} one recovers the maximum confidence in \cite{flatt2022} according to a noncontextual model.

For smaller rates $p_{\text{inc}}$, the support of the response functions corresponding to conclusive outcomes will shift to the support of the oposite states. In other words, we can write down these response functions as follows
\begin{align}
    &\xi_{0} (\lambda) = \begin{cases} a & \mbox{if } \lambda \in {\rm supp} [\mu_{0}(\lambda)]\cup {\rm supp} [\overline{\mu}_{1}(\lambda)] \\ 
    b & \mbox{if } \lambda \in {\rm supp} [\overline{\mu}_{0}(\lambda)]\cup{\rm supp} [\overline{\mu}_{1}(\lambda)] \\
    a-b & \mbox{if } \lambda \in  {\rm supp} [\mu_{0}(\lambda)]\cup {\rm supp} [\mu_{1}(\lambda)] \\
    0 & \text{otherwise}
    \end{cases} \\
    &\xi_{1} (\lambda) = \begin{cases} a & \mbox{if } \lambda \in {\rm supp} [\overline{\mu}_{0}(\lambda)]\cup {\rm supp} [\mu_{1}(\lambda)] \\ 
    b & \mbox{if } \lambda \in {\rm supp} [\overline{\mu}_{0}(\lambda)]\cup{\rm supp} [\overline{\mu}_{1}(\lambda)] \\
    a-b & \mbox{if } \lambda \in  {\rm supp} [\mu_{0}(\lambda)]\cup {\rm supp} [\mu_{1}(\lambda)] \\
    0 & \text{otherwise}
    \end{cases} \ .
\end{align}
Note that if $a=b$ we recover the response functions in \eqref{eq:respfuncbelow}. This allows us to rewrite the initial optimisation problem \eqref{eq:sdpbelow} in the following form
\begin{align}
    \underset{a,b}{\text{maximise}} & \quad \frac{1}{2}ar_{s}(1-c_{10}) \\
    \text{subject to} & \quad a-b \leq 2 \ , \quad  b\leq \frac{1}{2} \\
    & \quad p_{\text{inc}} = 1-a(1+r_{s}c_{10})+2br_{s}c_{10} \ .
\end{align}
The optimal values of the parameters $a$ and $b$ are
\begin{align}
    a = 1-\frac{p_{\text{inc}}}{1+r_{s}c_{10}} \quad b=\frac{1}{2} \ .
\end{align}
Then, one can write the success and error probabilities directly as follows:
\begin{align}
    p_{\text{suc}}^{\text{NC}} &= \frac{1+r_{s}}{2}\left(1-\frac{p_{\text{inc}}}{1+r_{s}c_{10}}\right) - \frac{r_{s}c_{10}}{2} \\ 
    p_{\text{err}}^{\text{NC}} &= \left(r_{s}c_{10}+\frac{1-r_{s}}{2}\right)\left(1-\frac{p_{\text{inc}}}{1+r_{s}c_{10}}\right)- \frac{r_{s}c_{10}}{2} \ .
\end{align}
One can use this result to obtain the maximum confidence for smaller inconclusive rates (i.e. for $p_{\text{inc}}\leq (1+rc_{10})/2$), which yields
\begin{align}
    \max C^{\text{NC}} = \frac{1}{2(1-p_{\text{inc}})}\left((1+r_{s})\left(1-\frac{p_{\text{inc}}}{1+r_{s}c_{10}}\right) - rc_{10}\right)
\end{align}
We can claim that this is the maximum confidence since it is also achieved by a measurement that simultaneously minimizes the error and maximises the success.

\end{widetext}

\end{document}